\documentclass[preprint,proceedings]{rmaa}

\usepackage{paralist}

\usepackage{psfrag,color}

\SetYear{2004}
\SetConfTitle{Compact Binaries in the Galaxy and Beyond}

\title{Binaries with Compact Components: Theoretical and Observational
Challenges} 

\author{
  Ph. Podsiadlowski,\altaffilmark{1} 
  E. Pfahl,\altaffilmark{2}
  and S. Rappaport\altaffilmark{3,}}

\altaffiltext{1}{Dep.\ of Astrophysics, University of Oxford, Oxford, 
OX1 3RH, UK.}
\altaffiltext{2}{Center for Astrophysics, Cambridge, MA 02138, USA.}
\altaffiltext{3}{Center for Space Research, M.I.T., Cambridge, MA 02139, USA.}

\shortauthor{Podsiadlowski, Rappaport, \& Pfahl}
\shorttitle{Binaries with Compact Components}

\fulladdresses{}
\listofauthors{Ph. Podsiadlowski, S. Rappaport, \& E. Pfahl}
\indexauthor{Podsiadlowski, Ph.}
\indexauthor{Rappaport, S.}
\indexauthor{Pfahl, E.}

\abstract{We report on recent progress in our theoretical
understanding of X-ray binaries, which has largely been driven by new
observations, and illustrate the interplay between theory and
observations considering as examples intermediate-mass X-ray
binaries, irradiation-driven evolution, ultraluminous X-ray sources
and neutron stars with low-velocity kicks.}
\addkeyword{Stars: neutron}
\addkeyword{Binaries: close}
\addkeyword{Black holes}
\addkeyword{X-rays: stars}

\begin{document}
\maketitle

\section{Introduction}
\label{sec:intro}
Our general understanding of binaries with compact components, in
particular those containing neutron stars and black holes still has serious
gaps, and theoretical progress is often driven by new observational
discoveries. Observations not only help to guide theorists, but also
provide important constraints that a successful theory has to satisfy.
In this contribution, we discuss the interplay between theory and observations
using several selected topics, including both neutron-star and black-hole
binaries.

\section{The Importance of Intermediate-Mass X-ray Binaries}
One of the major recent developments in the field of X-ray binary
research has been the realization that X-ray binaries with
intermediate-mass companion stars (IMXBs) are much more important than
believed previously. Indeed, IMXBs provide a particularly good example
that illustrate the interplay between theory and observations. 
The {\em observations} by Casares et al.\ (1998) showed that the
companion of the X-ray binary Cygnus X-2, formerly classified as a
low-mass X-ray binary (LMXB), was far too luminous and far too hot to
be consistent with a sub-giant in a 10-d orbit. The {\em theoretical}
resolution of this surprising observation (King \& Ritter 1999;
Podsiadlowski \& Rappaport 2000) was that the system must have
originated from an IMXB rather than an LMXB, where the mass of the
companion star must originally have been around 3.5\,$M_{\odot}$ (also
see Tauris et al.\ 2000; Kolb et al.\ 2000). However, this implies
that the system must have survived as a binary despite an extremely
high mass-transfer rate ($\sim 10^{-5}\,M_{\odot}\,{\rm yr}^{-1}$) --
several orders of magnitude above the Eddington accretion rate --
ejecting most of the transferred mass. How can a system eject all of
this mass? Again, {\em observations} may provide the essential
clues. Radio observations of the relativistic jet system SS 433
(Blundell et al.\ 2001) and possibly of Cygnus X-3 (Miller-Jones et
al.\ 2004, in preparation) show that most of the transferred mass in
these systems is lost in an equatorial, disk-like outflow.

Podsiadlowski, Rappaport \& Pfahl (2002) and Pfahl, Rappaport \&
Podsiadlowski (2003) have systematically investigated {\em
theoretically} the role of IMXBs and found, not surprisingly, that
IMXBs are much easier to form than traditional LMXBs, since these
systems can more easily survive as binaries both a common-envelope
phase and the supernova in which the neutron star is formed. After
the initial high mass-transfer phase, IMXBs are almost
indistinguishable from LMXBs; but since they have much higher
birthrates, Pfahl et al.\ (2003) predict that 80\,--\,95\,\% of all
L/IMXBs in fact originate from IMXBs.

How can {\em observations} help to confirm this prediction? Pfahl et
al.\ (2002a) proposed that the thousands of weak X-ray sources in the
Galactic center region, discovered in large numbers with {\it Chandra}
(Wang et al.\ 2002) are in fact the progenitors of IMXBs and HMXBs,
where a neutron star accretes matter from the wind of an
intermediate-mass companion before the latter fills its Roche lobe
(for alternative suggestions, see Willems \& Kolb 2003; Belczynski \&
Taam 2003). Bandyopadhyay et al.\ (2004) have obtained VLT
observations to look for infrared counterparts of some 70 of these
weak sources. These observations may already provide an important test
of this prediction.

Another {\em observational\/} test to distinguish between LMXBs and
IMXBs is to look for chemical anomalies. Many of the descendants of
IMXBs should be helium-rich and show evidence for CNO-processing. Such
anomalies may manifest themselves directly spectroscopically or
indirectly through their effects on X-ray bursts (Cumming 2003).

\section{Problems with the Standard Model and Irradiation-Driven
Evolution}

Pfahl et al.\ (2003) performed the first binary population synthesis
study of L/IMXBs using realistic binary evolution models. One of their
main conclusions was that the standard model for L/IMXBs failed to reproduce
some of the main features of the observed population. The two most
significant failures are: (1) the overproduction of L/IMXBs by a factor
of 10\,--\,100 (though consistent with the birthrate of binary millisecond
pulsars), and (2) the luminosity distribution, where the theoretical
distribution neither produces enough luminous L/IMXBs (with
$L_{\rm X}> 10^{37}\,{\rm ergs\,s}^{-1}$) nor reproduces the observed
correlation between X-ray luminosity and orbital period (Podsiadlowski
et al.\ 2002).

One major omission in the standard model is that it does not take into
account the strong X-ray irradiation of the secondary which can
fundamentally change the evolution of the system by either driving a
wind from the secondary (Ruderman et al.\ 1989) or by driving
expansion of the secondary (Podsiadlowski 1991).  Even a modest
expansion of the secondary ($\sim 10\,$\%) can drive mass-transfer
cycles (Hameury et al.\ 1993) where the mass-transfer rate $\dot{M}$
is larger than the rate without irradiation effects by a factor $\ga 10$,
which at the same time shortens the X-ray active lifetime by a
proportionate amount. Pfahl et al.\ (2003) demonstrated that the
inclusion of such mass-transfer cycles could potentially solve both of
the major problems mentioned above, by increasing the typical observed
X-ray luminosity by a factor of 10 or more and at the same time
eliminating the L/IMXB overproduction problem, but still producing enough
binary millisecond pulsars.

At the present time, the effects of irradiation on the secondary are
still very poorly understood. Phillips \& Podsiadlowski (2002) have
shown that the external irradiation can dramatically distort the shape
of the companion which has important implications for modelling
ellipsoidal lightcurves and determining radial-velocity curves of the
secondary. One of the key uncertainties is how much energy is
transported from the irradiated side to the back side by
irradiation-driven circulation.  Even the transport of only 1\,\% of
the intercepted irradiation energy can have drastic effects on the
appearance and the further evolution of the secondary. To help answer
these questions, Beer has developed a custom-designed 3-d stellar
hydrodynamics code to study the irradiation-induced circulation
(initially using a polytropic equation of state, which is now being
extended to include a thermodynamic equation; Beer \& Podsiadlowski
2002a,b). Some of his preliminary results show that the circulation
velocities are a significant fraction of the sound speed and that a
substantial amount of energy is transported to the backside in the
form of kinetic energy (rather than thermal energy) where it is
thermalized and raises the temperature by more than 1000\,K in the
case of an LMXB companion.

Again, observations will play in essential role in constraining the
theoretical models (in particular the turbulent viscosity in the outer
shear layer). These constraints may involve ellipsoidal light curves,
phase-dependent spectral variations and distortions of radial-velocity
curves. Indeed many of these effects have already been observed in a
number of systems (e.g.  HZ Her/Her X-1, Cyg X-2, Nova Sco, AA Dor).

\section{Ultraluminous X-Ray Binaries}

Ultraluminous X-ray sources (ULXs) are luminous X-ray sources outside
the nuclei of external galaxies, typically defined to have an X-ray
luminosity larger than $10^{39}\,{\rm ergs\,s}^{-1}$.  They were
originally discovered by {\it Einstein} (Fabbiano 1989) and have been
found in large numbers by {\it ROSAT} and most recently {\it Chandra}.
While it had been suggested (e.g. Colbert \& Mushotzky 1999) that
these may contain intermediate-mass black holes of
$10^2$\,--\,$10^4\,M_{\odot}$, it now seems more likely that at least
the majority form the luminous tail of the stellar-mass black-hole
binary distribution (e.g. King et al.\ 2001, 2004).

\begin{figure}[h]
\includegraphics[width=4.5cm,angle=-90]{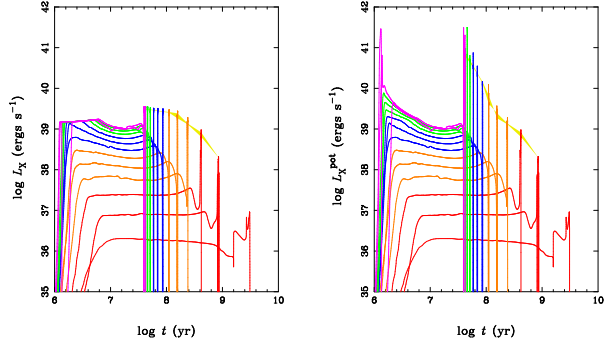}
\caption{X-ray luminosity, assuming Eddington-limited accretion,
(left) and {\em potential} X-ray luminosity, assuming non-Eddington
limited accretion, (right) for binary sequences containing a black
hole with an initial mass of 10\,$M_{\odot}$ and initially unevolved
secondaries from 2 to 17$\,M_{\odot}$ (roughly right to left; bottom
to top). (From Podsiadlowski et al.\ 2003)}
 \end{figure}

Podsiadlowski, Rappaport \& Han (2003) performed a systematic study of
the formation and the evolution of black-hole binaries using realistic
binary evolution calculations and found that indeed their models were
consistent with the observed luminosity function and the typical
number in a galaxy (of order one to a few). Figure~1 shows the X-ray
luminosity (left) and potential X-ray luminosity (right) as a function
of time for a sequence of binary models. The potential X-ray
luminosity is the luminosity of a system assuming that accretion is
not Eddington limited and that all the mass transferred from the
companion can be accreted, radiating at the appropriate accretion
efficiency. As the right panel shows, many of the more massive systems
have two phases in which the potential X-ray luminosities are in excess of
$10^{39}\,{\rm ergs\,s}^{-1}$, where these systems might appear as
ULXs; in an initial phase where mass-transfer occurs on a thermal
timescale and a later phase when the secondary evolves up the giant
branch, where the evolution is driven by hydrogen shell-burning.
Note, in particular, that the systems spend substantially more time
in the shell-burning phase than in the initial thermal timescale
phase. Indeed, GRS 1915+105, which is the only known Galactic ULX, is
well explained by these models.

In order for these systems to be ULXs requires a luminosity in excess
of the Eddington limit, typically by a factor of a few and less than a
factor 20 for even the most luminous systems, where this requirement
is further reduced if moderate geometrical beaming (King et
al.\ 2001) is important. We note that super-Eddington luminosities are
commonly observed in a number of neutron-star X-ray binaries,
presumably because the accretion flow is funnelled towards the poles
of the neutron star by magnetic fields. While this mechanism is not
applicable to black-hole binaries, Begelman (2002) has argued that
such super-Eddington luminosities can be explained in
radiation-pressure dominated, magnetic disks.

\section{A Dichotomous Kick-Scenario for Neutron-Star Kicks}

\begin{figure}[!t]
\centering
  \includegraphics[width=5.5cm]{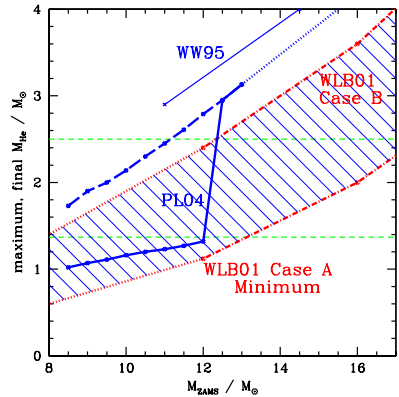}
  \caption{Final mass (thick solid line) and maximum mass (thick
dashed line) of the helium core in single stars as a function of
initial mass according to Poelarends \& Langer (2004, PL04),
extrapolated for initial masses above 12.5$\,M_{\odot}$ (the final
helium core masses from the calculations of Woosley \& Weaver [1995]
are indicated by a thin solid line). The hatched region shows the
final helium core masses expected in close binaries (from
the calculations of Wellstein et al.\ 2001 [WLB01]).  The light dashed
horizontal lines give the range for the final helium core mass for
which the star may experience an electron-capture supernova.  Note
that the parameter range for which this may occur for a single star is
very small.}
\end{figure}

It has long been established that neutron stars receive a kick with a
median velocity larger than $200\,{\rm km\,s}^{-1}$. Since such a
velocity is a factor of 5 to 10 larger than the central escape
velocity of even a massive globular cluster (GC), this implies that
most neutron stars forming from single stars should be ejected from 
clusters. On the other hand, rich GCs contain as many as $\sim 1000$
neutron stars (i.e.\ 10\,--\,20\,\% of the neutron stars formed).
This problem is known as the {\em neutron-star retention problem} (for
a detailed review see Pfahl et al.\ 2002b).

The retention problem is dramatically reduced if most neutron stars
are born in massive binaries, since in this case the momentum imparted
to the neutron star is shared with a companion star, leading to much
lower systemic velocities and making it much easier to retain the
system (Brandt \& Podsiadlowski 1995). As Pfahl et al.\ (2002b) have
shown this effect dramatically increases the number of neutron stars
that can be retained, although it may not be sufficient to explain the
observed numbers, unless globular clusters were initially much more
massive (Drukier 1996).

The problem can be substantially reduced if a significant fraction of
neutron stars only receive a small kick (Pfahl et al.\ 2002b). Indeed,
there is strong evidence that some neutron stars must receive rather
small kicks at birth from a newly established class of high-mass X-ray
binaries, which are relatively wide but have very low eccentricities
(Pfahl et al.\ 2002c). The prototype system is X
Per with an orbital period of 250\,d and an eccentricity of $\sim
0.10$.  Since the system is too wide for tidal effects to be
important, this requires that the neutron star can only have received a
moderate natal kick.

Pfahl et al.\ (2002c) and Podsiadlowski et al.\ (2004) speculated that
whether a neutron star receives a large or a small kick depends on
whether the progenitor was single or a member of a close binary, where
it lost its envelope soon after the main-sequence phase (i.e. in case
B mass transfer). As is not widely known, the evolution of the core of
a massive star and its final pre-supernova structure differs
substantially between single stars and stars in close binaries.
Massive stars that lose their envelopes in case B mass transfer
develop much smaller helium cores and ultimately smaller iron cores
(see Brown et al.\ 1999). Moreover, as the most up-to-date stellar
evolution calculations have shown (see Podsiadlowski et al.\ 2004),
massive stars in the range of 8\,--\,11\,$M_{\odot}$ that have lost
their hydrogen-rich envelopes before ascending the asymptotic giant
branch do not experience a second dredge-up phase, which would
dramatically reduce the mass of the helium core (see Fig.~2). This
suggests that single stars in this mass range most likely end their
evolution as ONeMg white dwarfs rather than in a supernova, while
stars that have lost their envelopes can explode in a supernova, most
likely an electron-capture supernova.  Podsiadlowski et al.\ (2004)
speculated that the core collapse in a small iron core or an
electron-capture supernova leads to a fast (prompt) explosion
where the instabilities that produce large kicks in more
massive cores do not have time to grow.  This suggests a dichotomous
scenario for neutron-star kicks, where stars in close binaries,
producing small pre-supernova cores, lead to fast (prompt) supernova
explosions with low kicks, while stars with more massive cores lead to
slow explosions with a standard high kick.
Indeed this scenario has recently received strong, theoretical support
from the core-collapse calculations by Scheck et al.\ (2004), which
show that the collapse of a small core produces a fast explosion with
a small kick, while the collapse of a massive core leads to a slow
explosion where convection-driven instabilities have time to grow and
produce large supernova kicks (but also see Fryer \& Warren 2004).


\begin{thebibliography}{}

\bibitem[(2999)]{} Bandyopadhyay, R.M., et al.\ 2004, these proceedings

\bibitem[(2999)]{} Beer, M.E.,  Podsiadlowski, Ph.\  2002a, MNRAS, 335, 358

\bibitem[(2999)]{} \ibidrule{} 2002b, in Tout,
C.A.  Van Hamme, W., eds, Exotic Stars as Challenges to Evolution,
ASP Conf.\ Proc., Vol.\ 279 (ASP, San Francisco), p.\ 253

\bibitem[(2999)]{} Begelman, M.C. 2002, ApJ, 568, 97

\bibitem[(2999)]{} Belczynski, K.,  Taam, R.E. 2004, ApJ, submitted
(astro-ph/0311287)

\bibitem[(2999)]{} Blundell, K.M. et al.\ 
2001, ApJL, 562, L79

\bibitem[Brandt  Podsiadlowski(1995)]{Brandt1995}
Brandt, W.N.,  Podsiadlowski, Ph.\ 1995, MNRAS, 274, 461

\bibitem[Brown, Lee  Bethe(1999)]{Brown1999}
Brown, G.E., Lee, C.-H., Bethe, H.A. 1999, NewA, 4, 313

\bibitem[(2999)]{} Casares, J., Charles, P.,  Kuulkers, E. 1998, ApJ, 493,
L39

\bibitem[(2999)]{} Colbert, E.J.M., Mushotzky, R.F. 1999, ApJ, 519, 89
\bibitem[(2999)]{} Cumming, A. 2003, ApJ, 595, 1077

\bibitem[Drukier(1996)]{Drukier1996} Drukier, G.A. 1996, MNRAS, 280, 498

\bibitem[(2999)]{} Fabbiano, G. 1989, ARA\&A, 27, 87


\bibitem[(2999)]{} Fryer, C.L.,  Warren, M.S. 2004, ApJ, in press (astro-ph/0309539)

\bibitem[(2999)]{} Hameury, J.M., King, A.R., Lasota, J.P.,  Raison,
 F. 1993, A\&A, 277, 81


\bibitem[(2999)]{} King, A.R. 2004, these proceedings

\bibitem[(2999)]{} King, A.R.,  Ritter, H. 1999, MNRAS, 309, 253

\bibitem[(2999)]{} King, A.R. et al.\ 2001, ApJ, 552, L109

\bibitem[(2999)]{} Kolb, U., Davies, M.B., King, A.,  Ritter, H. 2000, MNRAS, 
317, 438

\bibitem[(2999)]{} Phillips, S.N.,  Podsiadlowski, Ph.\ 2002, MNRAS, 337, 431

\bibitem[Pfahl, Rappaport,  Podsiadlowski(2002a)]{Pfahl2002a}
Pfahl, E., Rappaport, S.,  Podsiadlowski Ph. 2002a, ApJ, 571, L37 

\bibitem[Pfahl, Rappaport,  Podsiadlowski(2002b)]{Pfahl2002b}
\ibidrule{} 2002b, ApJ, 573, 283 

\bibitem[Pfahl, Rappaport,  Podsiadlowski(2003)]{Pfahl2003} 
\ibidrule{}2003, ApJ, 597, 1036 

\bibitem[Pfahl et al.(2002c)]{Pfahl2002c} 
Pfahl, E., Rappaport, S., Podsiadlowski, Ph.,  Spruit, H. 2002c,
ApJ, 574, 364  



\bibitem[(2999)]{} Podsiadlowski, Ph.\ 1991, Nat, 350, 136

\bibitem[(2999)]{} Podsiadlowski, Ph., Langer, N., Poelarends, A.J.T., Rappaport, S.,
Heger, A.,  Pfahl, E. 2004, ApJ, submitted (astro-ph/0309588)

\bibitem[(2999)]{} Podsiadlowski, Ph., Rappaport 2000, ApJ, 529, 946

\bibitem[Podsiadlowski, Rappaport,  Han(2003)]{Podsiadlowski2003}
Podsiadlowski, Ph., Rappaport, S.,  Han, Z. 2003, MNRAS, 341, 385

\bibitem[(2999)]{} Podsiadlowski, Ph., Rappaport, S., Pfahl. E. 2002, ApJ,
565, 1107

\bibitem[Poelarends  Langer(2004)]{Poelarends2004}
Poelarends, A.J.T.,  Langer N. 2004, in preparation

\bibitem[(2999)]{} Ruderman, M., Shaham, J.,  Tavani, M. 1989, ApJ, 336, 507

\bibitem[(2999)]{} Scheck, L., Plewa, T., Janka, H.-Th., Kifonidis, K., 
 M\"uller, E. 2004, PhRvL, 92, 1103
989, ApJ, 336, 507

\bibitem[(2999)]{}
Tauris, T.M., van den Heuvel, E.P.J.,  Savonije, G.J. 2000,
ApJ, 530, L93

\bibitem[(2999)]{} Wang, Q.D., Gotthelf, E.V.,  Lang, C.C. 2002, Nature,
415, 148

\bibitem[(2999)]{} Wellstein, S., Langer, N., Braun, H. 2001, A\&A, 369, 939

\bibitem[(2999)]{} Willems, B.,  Kolb, U. 2003, MNRAS, 343, 949

\bibitem[Woosley  Weaver(1995)]{Woosley1995b}
Woosley, S. E.,  Weaver, T. A. 1995, ApJS, 101, 181

\end{thebibliography}
\end{document}